# LOGIC MINING USING NEURAL NETWORKS

*W A T Wan Abdullah & + *Saratha Sathasivam

* Department of Physics, Universiti Malaya, 50603 Kuala Lumpur, Malaysia.
+ School of Mathematics, University Science Malaysia, 11800 Pulau Pinang, Malaysia.
E-mail: saratha@email.com

## Abstract

*Knowledge could be gained from experts, specialists in the area of interest, or it can be gained by induction from sets of data. Automatic induction of knowledge from data sets, usually stored in large databases, is called data mining. Data mining methods are important in the management of complex systems. There are many technologies available to data mining practitioners, including Artificial Neural Networks, Regression, and Decision Trees. Neural networks have been successfully applied in wide range of supervised and unsupervised learning applications. Neural network methods are not commonly used for data mining tasks, because they often produce incomprehensible models, and require long training times.*

*One way in which the collective properties of a neural network may be used to implement a computational task is by way of the concept of energy minimization. The Hopfield network is well-known example of such an approach. The Hopfield network is useful as content addressable memory or an analog computer for solving combinatorial-type optimization problems. Wan Abdullah [1] proposed a method of doing logic programming on a Hopfield neural network. Optimization of logical inconsistency is carried out by the network after the connection strengths are defined from the logic program; the network relaxes to neural states corresponding to a valid interpretation.*

*In this article, we describe how Hopfield network is able to induce logical rules from large database by using reverse analysis method: given the values of the connections of a network, we can hope to know what logical rules are entrenched in the database.*

## Key-words:
 Hopfield, Logic Programming, data mining, neural network

## 1.0 Introduction

The main focus of the data mining task is to gain insight into large collections of data. Often achieving this goal involves applying machine-learning methods to inductively construct models of the data at hand. Although neural network learning algorithms have been successfully applied in wide range of supervised and unsupervised learning applications, they have not often been applied in data mining settings, in which two fundamental considerations are the comprehensibility and speed issues which often are of prime importance in the data mining community.

Data mining is not merely automatic collecting of knowledge . Human-computer collaboration knowledge discovery is the interactive process between data miner and and computer. The aim is to ext ract novel, plausible, relevant and interesting knowledge from the database. We do not provide an introduction to data mining techniques in this paper, but instead refer the interested reader to one of the good book in the field [ 2 ].

Logic programming can be treated as a problem in combinatorial optimization. Therefore it can be carried out in a neural network to obtain the desired solution. Our objective is to find a set of interpretation (i.e., truth value assignments) for the atoms in the clauses which satisfy the clauses (which yields all the clauses true). We extended the work related to logic programming in neural network by introducing reverse analysis method. This method is capable to induce logical rules entrenched in a database. The knowledge obtained from the logical rules can be used to unearth relationship in data that may provide useful insights.

The rest of the paper organized as follows. In the next section we consider some theory of the Little- Hopfield Model. Section 3 discusses about logic programming focusing on Horn clauses. Follow by Section 4, where the logic of Hebbian learning will be discussed. Section 5 describes method for extracting rules from database: Reverse analysis method. Finally, Section 6 provides discussion and conclusion.

## 2.0 Little - Hopfield Model

In order to keep this paper self-contained we briefly review the Little Hopfield Model [3]. The Hopfield model [4] is a standard model for associative memory. The Little dynamics is asynchronous, with each neuron update their state deterministically. It consists of N formal neurons, each of which is described by an Ising variable $S_i(t)(i=1,2,....N)$. Neurons can be modeled as being binary, $V_i \in \{0,1\}$, obeying the dynamics $V_i \to q(h_i)$ where $h_i = \sum_j T_{ij}^{(2)} V_j + T_i^{(1)}$, $i$ and $j$ running over all neurons $N$, $T_{ij}^{(2)}$ is the synaptic strength from $j$ to neuron $i$, $-T_i$ is the threshold of neuron $I$, and $q$ is the step function.

Alternatively, they can be taken to be bipolar, $S_i \in \{-1,1\}$ with $S_i \to \text{sgn}(h_i)$ where $V_i$ is replaced by $S_i$ in $h_i$. In the following, we write mainly expression for the binary case; the corresponding ones for the bipolar case can deduced accordingly.

Restricting the connections to be symmetric and zero-diagonal, $T_{ij} = T_{ji}$, $T_{ii} = 0$, allows one to write Lyapunov energy function as

$$E = -\frac{1}{2}\sum_i\sum_j T_{ij}V_iV_j - \sum_i T_iV_i \quad (1)$$

which monotone decreases with the dynamics.

The two-connection model can be generalized to include higher order connections. This modifies the "field" to be

$$h_i = .... + \sum_j\sum_k T_{ijk}^{(3)}V_jV_k + \sum_j T_{ij}^{(2)}V_j + T_i^{(1)} \quad (2)$$

where "....." denotes still higher orders, and an energy function can be written as follows:

$$E = ..... - \frac{1}{3}\sum_i\sum_j\sum_k T_{ijk}V_iV_jV_k - \frac{1}{2}\sum_i\sum_j T_{ij}V_i V_j - \sum_i T_iV_i \quad (3)$$

provided that $T_{ijk} = T_{[ijk]}$ for $i,j,k$ distinct, which[...] denoting permutations in order, and $T_{ijk} = 0$ for any $i,j,k$ equal, and that similar symmetry requirements are satisfied for higher order connections. An updating rule reads

$$S_i(t+1) = \text{sgn}[h_i(t)] \quad (4)$$

## 3.0 Logic Programming

In the simple propositional case, logic clauses take the form

$$A_1, A_2,......,A_n \leftarrow B_1, B_2,....,B_m.$$

which says that ($A_1$ or $A_2$ or....or $A_n$) if ($B_1$ and $B_2$ and...and $B_n$);they are Horn clauses if $n=1$ and $m \geq 0$ : we can have rules e.g. $A \leftarrow B,C.$ saying $A \lor \neg(B \land C) = A \lor \bar{B} \lor \bar{C}$, and assertions e.g. $D \leftarrow .$ saying that $D$ is true.

A logic program consists of a set of Horn clause procedures and is activated by an initial goal statement. It is in the form of Conjunctive Normal Form (CNF) and contains one positive literal.

Basically, logic programming in Hopfield model [5] can be treated as a problem in combinatorial optimization. Therefore it can be carried out in a neural network to obtain the desired solution. Our objective is to find a set of interpretation (i.e., truth value assignments) for the atoms in the clauses which satisfy the clauses (which yields all the clauses true).

For an example, consider the logic program below:

$$P = A \leftarrow B,C$$
$$\leftarrow D \leftarrow B$$
$$\leftarrow C \leftarrow .$$

Given the goal
$$\leftarrow G$$

we require to show that $P \wedge \neg G$ is inconsistent in order to prove the goal. Alternatively, we require to find an interpretation for the Herbrand base of the problem which is consistent with $P$ (which yields $P$ true) and examine the truth of $G$ in such an interpretation. If we assign the values *1* to *true* and *0* to *false* then $\neg P = 0$ indicates a consistent interpretation while $\neg P = 1$ reveals that at least one of the clauses in the program is not satisfied. Therefore, looking for a consistent interpretation is a combinatorial (of assigning truth values to ground atoms) minimization of the inconsistency, the value of $\neg P$.

Translate all clauses and the negation of it into Boolean algebraic form:

$$P = (A \vee \neg B \vee \neg C) \wedge (D \vee \neg B) \wedge C$$
$$\neg P = (\neg A \wedge B \wedge C) \vee (\neg D \wedge B) \vee (\neg C)$$

From these, we may write a cost function which is minimized when all the clauses are satisfied as follow:

$$E_p = (1 - V_A)V_B V_C + (1 - V_D)V_B + (1 - V_C) \quad (5)$$

where the neurons $V_A$, as an example, represent the truth values of *A*. Notice that we have chosen the multiplication operation to represent the relationship "AND", and addition operation "OR".

The minimum value for $E_p$ is 0, corresponding to the fact that all the clauses are satisfied. The value for $E_p$ (which is an integer) is proportional to the number of clauses unsatisfied.

An energy function is defined as:

$$H = -\frac{1}{3}\sum_i \sum_j \sum_k T_{ijk} V_i V_j V_k - \frac{1}{2}\sum_i \sum_j T_{ij} V_i V_j - \sum_i J_i V_i \quad (6)$$

where the synaptic strength is completely symmetric with zeros in the diagonal planes.

By comparing (5) and (6), we obtained the connection strengths.

## 4.0 The Logic Of Hebbian Learning

Now we reproduce results in [5] which calculate the connection strengths using Hebb Rule [6].

For two-neuron connections, a Hebbian-like learning is given by

$$\Delta T_{ij} = a(2V_i - 1)(2V_j - 1) \quad (7)$$

(or, for bipolar neurons, $\Delta T_{ij} = a S_i S_j$), where $a$ is a learning rate. For connections of other orders, we can generalized this to

$$\Delta T_{ij....m} = a(2V_i - 1)(2V_j - 1)..(2V_n - 1) \quad (8)$$

Assume we have events $A, \bar{A}, B, \bar{B}, C, \bar{C},...$ occurring randomly but equally probably: $V_a$, etc. are randomly 0 or 1 with equal probabilities. As such, there would be no nett change in connection strengths, because $\Delta T_{ij...n}$ has equal probability of being positive as well as being negative.

Now say for example $\bar{D}$ does not occur. This would result in $\Delta T_D$ being positive, which is equivalent to, according to our analysis in the previous section, the assertion $D \leftarrow .$ being learnt. In this case, our system has learnt a rule which corresponds to only *D* occurring.

For the case of *C* occurring when *D* occurs, example *CD* occurs, $C\bar{D}$ occurs, $\bar{C}D$ does not occur, $\bar{C}\bar{D}$ occurs, there is a nett increase in $\Delta T_{[CD]}$ and a nett decrease of the same magnitude in $\Delta T_D$. This is equivalent to the rule $C \leftarrow D.$ being learnt, which the rule giving events like those is observed. However, there is also an increase in $\Delta T_C$ which means that

$C \leftarrow .$ has also been learnt .

To clarify further, let us look at the case where $A$ occurs if $B$ and $C$ both do. Then the following table summarizes what happens:

|  | $\Delta T_{[ABC]}$ | $\Delta T_{[AB]}$ | $\Delta T_{[AC]}$ | $\Delta T_{[BC]}$ | $\Delta T_A$ | $\Delta T_B$ | $\Delta T_C$ |
|---|---|---|---|---|---|---|---|
| $ABC$ occurs | + | + | + | + | + | + | + |
| $AB\bar{C}$ occurs | − | + | − | − | + | + | − |
| $A\bar{B}C$ occurs | − | − | + | − | + | − | + |
| $A\bar{B}\bar{C}$ occurs | + | − | − | + | + | − | − |
| $\bar{A}\bar{B}\bar{C}$ occurs | − | − | − | + | − | + | + |
| $\bar{A}\bar{B}C$ occurs | + | − | + | − | − | + | − |
| $\bar{A}\bar{B}\bar{C}$ occurs | + | + | − | − | − | − | + |
| $\bar{A}\bar{B}\bar{C}$ occurs | − | + | + | − | − | − | − |
| nett | + | + | + | - | + | - | - |
| factor | x 6(-1/3) | x 2(-1/2) | x 2(-1/2) | x 2(-1/2) | x 1(-1) | x 1(-1) | x 1(-1) |

The net change is multiplied by the number of terms in the energy giving the same contribution( the various permutations of the subscripts) and the factor associated with each term in the energy function. The system "correctly" learns $A \leftarrow B, C.$ but also the extra rules $A \leftarrow B., A \leftarrow C., A \leftarrow.$ and so on.

If we use bipolar neurons, the energy change to include the clause $C \leftarrow D.$ is

$$\Delta E = \frac{1}{4}(1 - S_c)(1 + S_D) \qquad (9)$$

Thus the events $\{CD, C\bar{D}, \bar{C}\bar{D}\}$ correctly give the corresponding change in energy without the spurious clause as with binary neurons. This may be expected as the change in variable is effectively an overall change in the neural threshold values.

For $A \leftarrow B, C.$ the change in energy with bipolar neurons is

$$\Delta E = \frac{1}{8}(1 - S_A)(1 + S_B)(1 + S_C) \qquad (10)$$

The collection of events {$ABC$, $AB\bar{C}$, $A\bar{B}C$, $A\bar{B}\bar{C}$, $\bar{A}BC$, $\bar{A}B\bar{C}$, $\bar{A}\bar{B}C$, $\bar{A}\bar{B}\bar{C}$} yields the learning of $A \leftarrow B, C.$ plus the extra energy term - $S_A S_B S_C$ which causes the system to have a liking for all $A$, $B$, and $C$ to be true.

## 5.0 Extracting Rules From Database

Companies have been collecting data for decades, building massive data warehouses in which to store it. Even though this data is available, very few companies have been able to realize the actual value stored in it. The question these companies are asking is how to extract this value?

So, in this paper we proposed a method known as reverse analysis to induce the logical rules entrenched in a database. These logical rules represent significant patterns or trends in the database that would otherwise go unrecognized.

In this section, we describe the implementation of our method regarding extracting rules from database.

i) Enumerate number of neurons and patterns in the database
*ii)* Extract the events from the database and represent in binary/bipolar pattern, where 0 indicates false state and 1 indicates true state (for bipolar -1 represent false state and 1 represent true state).
*iii)* Calculate the connection strengths for the events using Hebbian learning as indicated in Section 4.0.
*iv)* Capture nonzero values (connection strengths) for third order connection.
*v)* Reverse analysis been carried out to deduce the underlying logical rules.
*vi)* Logical rules represented in the form of Horn clauses.
*vii)* Calculate connection strengths for the extracted Horn clauses and deduct the value of the connection strengths from (iii).
*viii)* Repeat the similar steps for second-order and first-order connection.

Reverse analysis method has been tested in a small data set as shown in Table 1. The logical rules induced from the method seem to agree with the frequent observations.

Table 1: Customers daily purchased from a supermarket:

|       | Bread | Jam | Cheese | Sausage | Burger |
|-------|-------|-----|--------|---------|--------|
| Peter | v     |     |        |         |        |
| Sue   |       | v   | v      | v       |        |
| John  | v     |     |        |         | v      |
| Mary  |       | v   |        |         | v      |
| Anne  | v     |     | v      | v       |        |

From the reverse analysis method discussed in Section 5, we obtained the following rules:

Bread ← Jam, Cheese
Burger ← Sausage, Cheese

The logical rules induced, can help the departmental store in monitoring their stock according to the customers demand. Significant patterns or trends in the data set have been identified by using reverse analysis. The departmental store can apply the patterns to improve its sales process according to customers shopping trends. Futhermore, the knowledge obtained may suggest new initiatives and provide information that improves future decision making.

## 6.0 CONCLUSION

Data might be one of the most valuable assets if we know how to reveal valuable knowledge hidden in raw data. By doing reverse analysis: given the values of the connections of a network (obtained from the data set) , we can hope to know what logical rules are entrenched in it. Reverse analysis method can help us in revealing knowledge hidden in data and turn this knowledge into a crucial competitive advantage.

The reverse analysis method yields some limitation such as exists of extra terms (discussed in Section 4.0) and redundancy clauses due to interference effects. However, if we take the learning rate for three-neuron connections to be half that for two-neuron connections, this extra term is lost. Moreover, redundancies clauses have been proved does no effect the knowledge base [7].

The logical rules are obtained through frequent observation and these are not necessarily intrinsic in the object themselves. It is our hope that our reverse analysis method will serve to inspire some interesting applications of this method to challenging data mining tasks.